\documentclass[final]{siamltex}

\usepackage{epsfig}

\title{Finding two-dimensional peaks}

\author{Z.~K.~Silagadze
} 

\begin{document}
                                                                            
\maketitle 

\vspace*{-3mm}
\centerline
{\it Budker Institute of Nuclear Physics,  630 090, Novosibirsk, Russia}

\vspace*{2mm} 
\begin{abstract}
Two-dimensional generalization of the original peak finding algorithm 
suggested earlier is given. The ideology of the algorithm emerged from the
well known quantum mechanical tunneling property which enables small bodies
to penetrate through narrow potential barriers. We further merge this 
``quantum'' ideology with the philosophy of Particle Swarm Optimization to
get the global optimization algorithm which can be called Quantum Swarm 
Optimization. The functionality of the newborn algorithm is tested on some
benchmark optimization problems.  
\end{abstract}

\begin{keywords}
Numerical optimization, Global optimum, Quantum Swarm Optimization
\end{keywords}


\section{Introduction}
Some time ago we suggested a new algorithm for automatic photopeak location
in gamma-ray spectra from semiconductor and scintillator detectors \cite{1}.
The algorithm was inspired by quantum mechanical property of small balls to
penetrate through narrow barriers and find their way down to the potential
wall bottom even in the case of irregular potential shape.

In one dimensional case the idea was realized by means of finite Markov 
chain and its invariant distribution \cite{1}. States of this Markov chain
correspond to channels of the original histogram. The only nonzero 
transition probabilities are those which connect a given state to its 
closest left and right neighbor states. Therefore the transition probability
matrix for our Markov chain has the form
$$ P=\left ( \begin{array}{cccccccc}
0 & 1 & 0 & 0 & 0 & \cdot & \cdot & \cdot \\
P_{21} & 0 & P_{23} & 0 & 0 & \cdot & \cdot & \cdot \\
0 & P_{32} & 0 & P_{34} & 0 & \cdot & \cdot & \cdot \\
\cdot & \cdot & \cdot & \cdot & \cdot & \cdot & \cdot & \cdot \\
0 & \cdot & \cdot & \cdot & \cdot & 0 & 1 & 0
\end{array} \right ) \hspace*{2mm} . $$
As for the transition probabilities, the following expressions were used
\begin{equation}
P_{i,i \pm 1} =\frac{Q_{i,i \pm 1}}{Q_{i,i-1}+Q_{i,i+1}}
\label{eq1} \end{equation}
with 
\begin{equation}
Q_{i,i \pm 1}= \sum_{k=1}^{m} \exp { \left [ \frac{N_{i \pm k}
-N_i}{\sqrt{N_{i \pm k}+N_i}} \right ] } \hspace*{2mm} . 
\label{eq2} \end{equation}
The number $m$ is a parameter of the model which mimics the (inverse) mass
of the quantum ball and therefore allows to govern its penetrating ability.

The invariant distribution for the above described Markov 
chain can be given by a simple analytic formula \cite{2}
$$
u_2=\frac{P_{12}}{P_{21}}u_1 \; , \; u_3=\frac{P_{12}P_{23}}{P_{32}
P_{21}}u_1 \; , \; \cdots \; , \; u_n=\frac{P_{12}P_{23} \cdots
P_{n-1,n}}{P_{n,n-1}P_{n-1,n-2} \cdots P_{21}}u_1 \hspace*{2mm} ,
$$
\noindent
where $u_1$ is defined from the normalization condition
$$
\sum_{i=1}^n u_i = 1 \hspace*{2mm} .
$$

Local maximums in the original spectrum are translated into the very sharp
peaks in the invariant distribution and therefore their location is 
facilitated. 

The algorithm proved helpful in uniformity studies of NaJ(Tl) crystals for 
the SND detector \cite{3}. Another application of this ``peak amplifier'', to 
refine the amplitude fit method in ATLAS $B_s$-mixing studies, was described 
in \cite{4}. In this paper we will try to extend the method also in the 
two-dimensional case.

\section{Two-dimensional generalization}
The following two-dimensional generalization seems straightforward. For 
two-dimensional $n\times n$ histograms the corresponding Markov chain states 
will also form a two-dimensional array $(i,j)$. Let $u^{(k)}_{ij}$ be 
a probability for the state $(i,j)$ to be occupied after $k$-steps of the 
Markov process. Then  
$$u^{(k+1)}_{lm}=\sum_{i,j=1}^n P_{ij;lm} u^{(k)}_{ij},$$
where $P_{ij;lm}$ is a transition probability from the state $(i,j)$ to the
state $(l,m)$. We will assume that the only nonzero transition probabilities 
are those which connect a given state to its closest left, right, up or down 
neighbor states. Then the generalization of equations (\ref{eq1}) and
(\ref{eq2}) is almost obvious. Namely, for the transition probabilities we
will take
 \begin{eqnarray} 
P_{ij;i,j \pm 1} & = & \frac{Q_{ij;i,j \pm 1}}
{Q_{ij;i,j-1}+Q_{ij;i,j+1}+Q_{ij;i-1,j}+Q_{ij;i+1,j}},\nonumber \\
P_{ij;i \pm 1,j} & = & \frac{Q_{ij;i \pm 1,j}}
{Q_{ij;i,j-1}+Q_{ij;i,j+1}+Q_{ij;i-1,j}+Q_{ij;i+1,j}},
\label{eq3} \end{eqnarray}
with 
\begin{eqnarray}
Q_{ij;i,j \pm 1} & = & \sum_{k=1}^{m} \sum_{l=-k}^{k}\exp { \left [ 
\frac{N_{i+l,j\pm k}-N_{ij}}{\sqrt{N_{i+l,j\pm k}+N_{ij}}} \right ] }
,\nonumber \\
Q_{ij;i \pm 1,j} & = & \sum_{k=1}^{m} \sum_{l=-k}^{k}\exp { \left [ 
\frac{N_{i \pm k,j+l}-N_{ij}}{\sqrt{N_{i \pm k,j+l}+N_{ij}}} \right ] 
}.
\label{eq4} \end{eqnarray}

We are interested in invariant distribution $u_{ij}$ for this Markov chain,
such that
$$\sum_{i,j=1}^n P_{ij;lm} u_{ij}=u_{lm}.$$
Unfortunately, unlike to the one-dimensional case, this invariant distribution
can not be given by a simple analytic formula. But there is a way out: having
at hand the transition probabilities $P_{ij;lm}$, we can simulate the 
corresponding Markov process starting with some initial distribution 
$u^{(0)}_{ij}$. Then after a sufficiently large number of iterations we will 
end with almost invariant distribution irrespective to the initial choice of 
$u^{(0)}_{ij}$. For example, in the role of $u^{(0)}_{ij}$ one can take the 
uniform distribution:
$$u^{(0)}_{ij}=\frac{1}{n^2}.$$

\begin{figure}[htbp]
\includegraphics[width=0.47\textwidth]{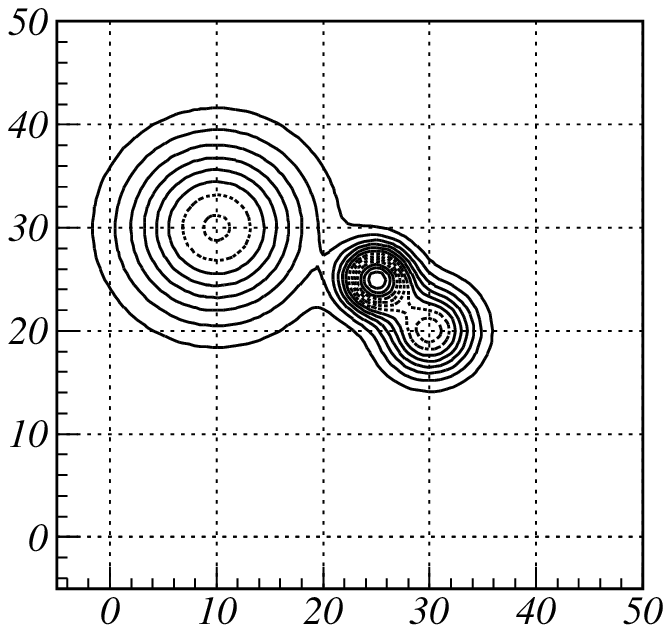}
\hfill
\includegraphics[width=0.51\textwidth]{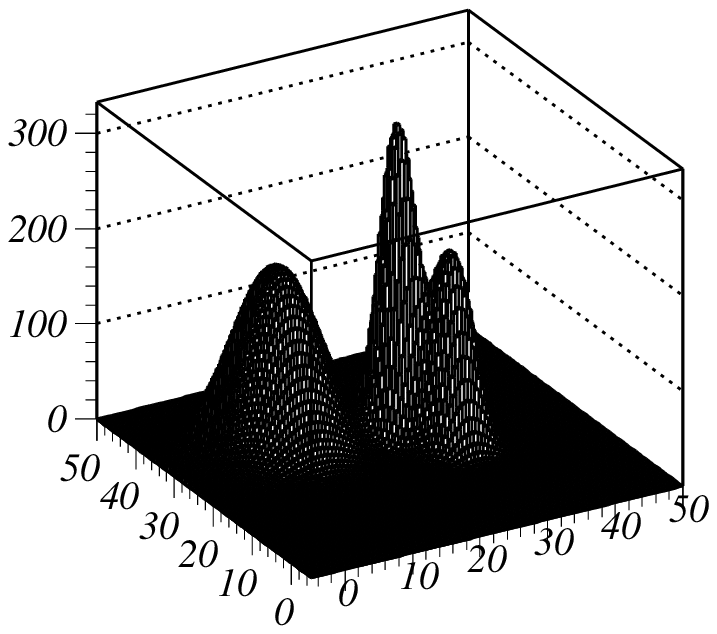}
\hfill
\includegraphics[width=0.47\textwidth]{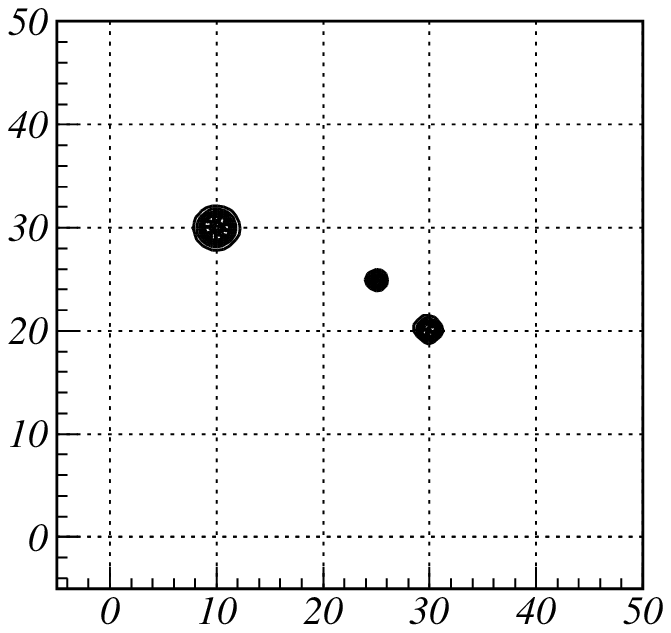}
\hfill
\includegraphics[width=0.51\textwidth]{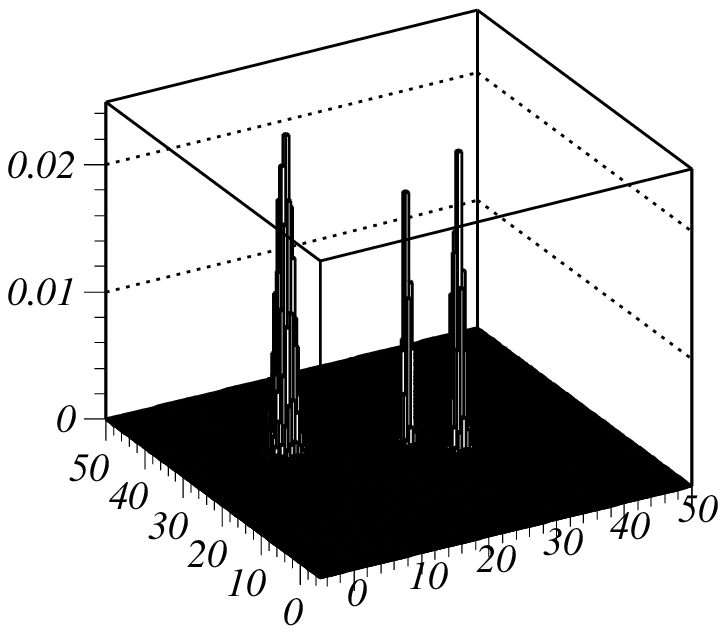}
\hfill
\includegraphics[width=0.47\textwidth]{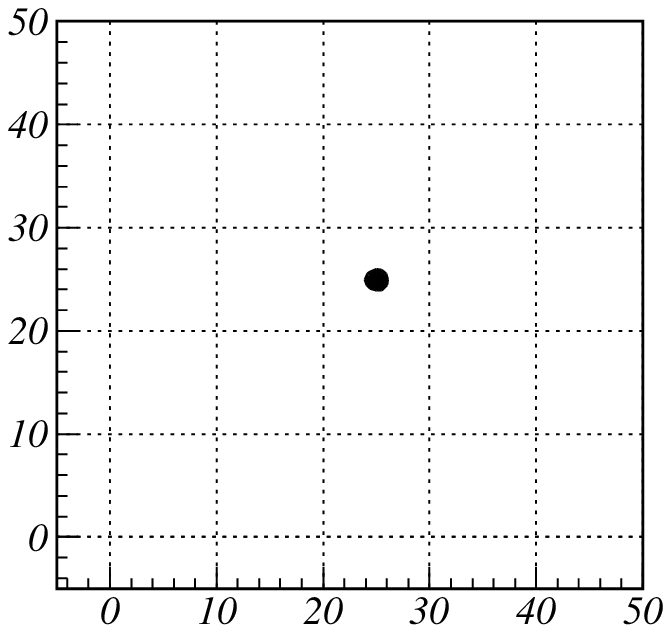}
\hfill
\includegraphics[width=0.51\textwidth]{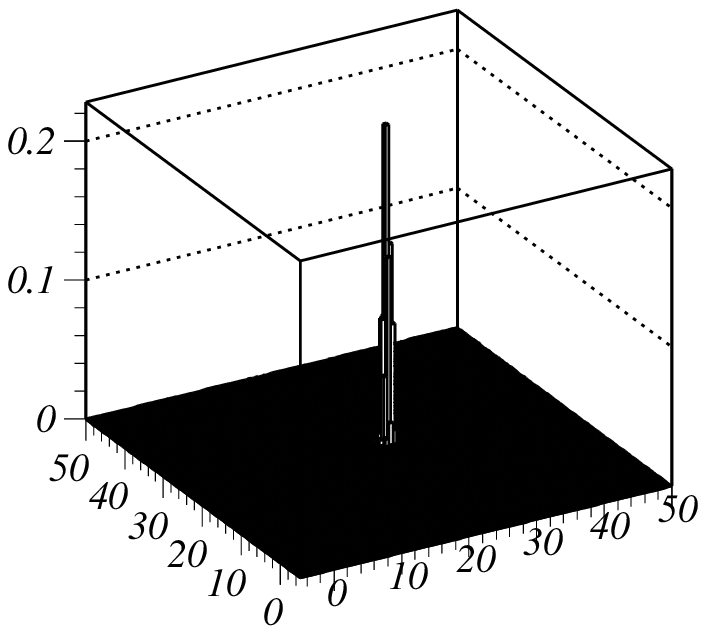}
\hfill
\caption{The upper histograms represent the initial data in the
contour and lego formats respectively. The middle histograms show the 
corresponding probability distribution after 258 iterations for the 
penetrating ability $m=3$. The lower histograms represent the invariant 
distribution for the penetrating ability $m=30$.}
\label{fig1}
\end{figure}

For practical realization of the algorithm, it is desirable to have precise 
meaning of words ``sufficiently large number of iterations''. In our first 
tests the following stopping criterion was used. One stops at $k$-th iteration 
if the averaged relative difference between $u^{(k)}_{ij}$ and 
$u^{(k-1)}_{ij}$ probability distributions  is less than the desired 
accuracy $\epsilon$:
\begin{equation}
\sum_{u^{(k)}_{ij}\ne 0} 2\;\frac{|u^{(k)}_{ij}-u^{(k-1)}_{ij}|}{u^{(k)}_{ij}
+u^{(k-1)}_{ij}}\; u^{(k)}_{ij} < \epsilon \, .
\label{eq5} \end{equation}

The performance of the algorithm is illustrated by Fig.\ref{fig1} for 
a $100\times 100$ histogram representing three overlapping Gaussians with 
different widths. As expected, it works much like to its one-dimensional 
cousin: the invariant probability distribution shows sharp peaks at locations
where the initial data has broad enough local maximums. Note that in this 
concrete example the one iteration variability $\epsilon=10^{-3}$ was reached 
after 258 iterations for $m=3$ and after 113 iterations for $m=30$.

Convergence to the invariant distribution can be slow. In the example given 
above by Fig.\ref{fig1} the convergence is indeed slow for small penetrating 
abilities. If we continue iterations for $m=3$ further, the side peaks will
slowly decay in favor of the main peak corresponding to the global maximum.
In the case of $m=3$ it takes too much  iterations to reach the invariant 
distribution. However, as Fig.\ref{fig1} indicates, the remarkable property 
to develop sharp peaks at locations of local maximums of the initial histogram 
is already revealed by $u^{(k)}_{ij}$ when number of iterations $k$ is of the 
order of 300.

\begin{figure}[htbp]
\includegraphics[width=0.47\textwidth]{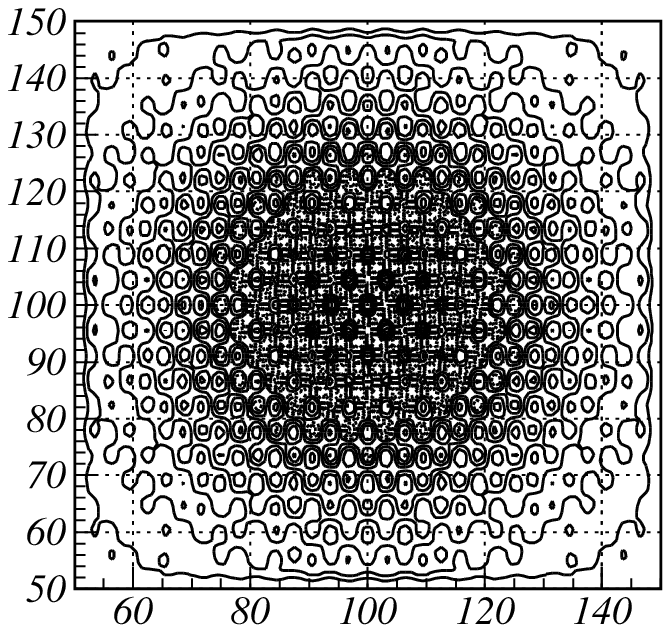}
\hfill
\includegraphics[width=0.51\textwidth]{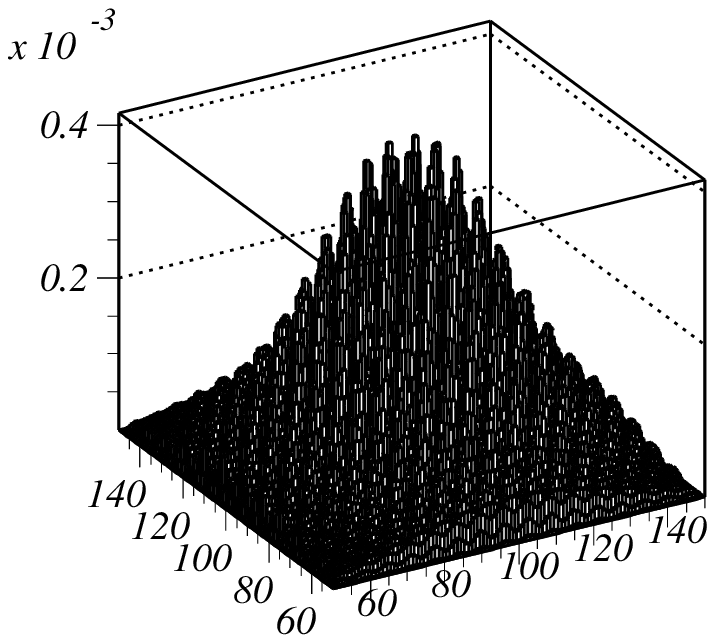}
\hfill
\includegraphics[width=0.47\textwidth]{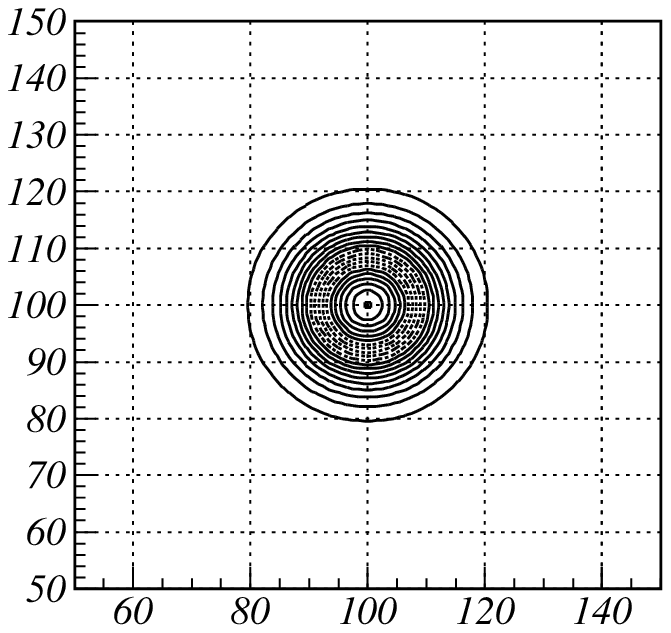}
\hfill
\includegraphics[width=0.51\textwidth]{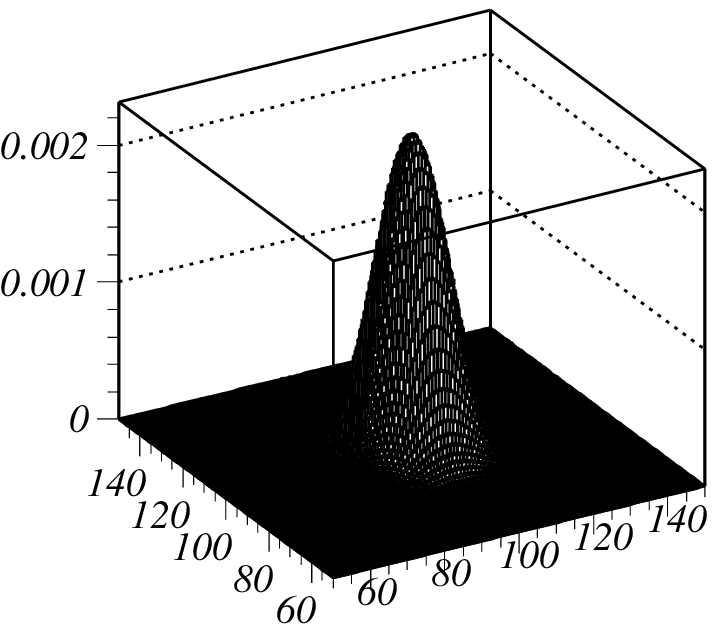}
\hfill
\caption{The probability distribution
for the Griewank function. Upper histograms for $m=3$, lower histograms for
$m=30$.}
\label{fig2}
\end{figure}

One can make the algorithm to emphasize minimums, not maximums, by just 
reversing signs in the exponents:
$$\exp { \left [\frac{N_{lm}-N_{ij}}{\sqrt{N_{lm}+N_{ij}}}\right ] } 
\longrightarrow
\exp { \left [-\frac{N_{lm}-N_{ij}}{\sqrt{N_{lm}+N_{ij}}}\right ] }.$$ 
This is illustrated by Fig.\ref{fig2}. Here the initial histogram is generated
by using a variant of the Griewank function \cite{5}
\begin{equation}
F(x,y)=\frac{ (x-100)^2+(y-100)^2}{4000}-\cos {(x-100)}
\cos{\frac{y-100}{\sqrt{2}}}+1.
\label{eq6} \end{equation}
This function has the global minimum at a point $x=100,\;y=100$ and in the
histogramed interval $50\le x \le 150,\; 50\le y \le 150$ exhibits nearly
thousand local minimums. Many of them are still visible in the probability 
distribution for penetrating ability $m=3$. But for $m=30$ only one peak, 
corresponding to the global minimum, remains. 

\section{Quantum swarm optimization}
The discussion above was focused on two-dimensional histograms, while in 
practice more common problem is finding global optimums of nonlinear 
functions. The algorithm in the form discussed so far is not suitable for 
this latter problem. However it is possible to merge its ideology with the 
one of particle swarm optimization \cite{6,7,8} to get a workable tool.
 
The particle swarm optimization was inspired by intriguing ability of bird 
flocks to find spots with food, even though birds in the flock had no previous 
knowledge of their location and appearance. ``This algorithm belongs 
ideologically to that philosophical school that allows wisdom to
emerge rather than trying to impose it, that emulates nature rather than 
trying to control it, and that seeks to make things simpler rather than more 
complex'' \cite{6}. This charming philosophy is indeed very attractive. So
we attempted to develop quantum swarm optimization - when each particle in
the swarm mimics the quantum behavior.

The algorithm that emerged goes as follows:
\begin{itemize}
\item initialize a swarm of $n_p$ particles at random positions in the search
space $x_{min}\le x \le x_{max},\;\; y_{min}\le y \le y_{max}$.
\item find a particle $i_b$ in the swarm with the best position $(x_b,y_b)$, 
such that the function $F(x,y)$ under investigation has the most optimal value 
for the swarm at $(x_b,y_b)$.
\item for each particle in the swarm find the distance from the best position
$d=\sqrt{(x-x_b)^2+(y-y_b)^2}$. For the best particle instead 
take the maximal value of these distances from the previous iteration (or 
for the first iteration take
$d=\sqrt{(x_{max}-x_{min})^2+(y_{max}-y_{min})^2}$ ).
\item generate a random number $r$ uniformly distributed in the interval
$0\le r \le 1$ and find the random step $h=rd$.
\item check for a better optimum the closest left, right, up and down
neighbor states with the step $h$. If the result is positive, change $i_b$ 
and $(x_b,y_b)$ respectively. Otherwise
\item move the particle to left, right, up or down by the step $h$ according
to the corresponding probabilities of such jumps:
\begin{eqnarray}
p_L=\frac{q_L}{q_L+q_R+q_U+q_D},\;\;\; 
p_R=\frac{q_R}{q_L+q_R+q_U+q_D}, \nonumber 
\\ p_U=\frac{q_U}{q_L+q_R+q_U+q_D},\;\;\; p_D=\frac{q_D}{q_L+q_R+q_U+q_D}, 
\label{eq7} \end{eqnarray}
where
\begin{eqnarray}
q_L=\sum_{y^\prime=y_u,y,y_d}\exp{\left ( I_s \frac{F(x_d,y^\prime)-
F(x,y)}{h} \right )}, \nonumber \\
q_R=\sum_{y^\prime=y_u,y,y_d}\exp{\left ( I_s \frac{F(x_u,y^\prime)-
F(x,y)}{h} \right )}, \nonumber \\
q_D=\sum_{x^\prime=x_u,x,x_d}\exp{\left ( I_s \frac{F(x^\prime,y_d)-
F(x,y)}{h} \right )}, \\
q_U=\sum_{x^\prime=x_u,x,x_d}\exp{\left ( I_s \frac{F(x^\prime,y_u)-
F(x,y)}{h} \right )}, \nonumber 
\label{eq8} \end{eqnarray}
and 
\begin{eqnarray} &&
x_u=\min{(x+h,x_{max})}, \;\; x_d=\max{(x-h,x_{min})}, \nonumber \\ &&
y_u=\min{(y+h,y_{max})}, \;\; y_d=\max{(y-h,y_{min})}.
\label{eq9} \end{eqnarray}
At last  $I_s=1$, if optimization means to find the global maximum, and 
$I_s=-1$, if the global minimum is searched.
\item do not stick at walls. If the particle is at the boundary of the search
space, it jumps away from the wall with the probability equaled one (that is 
the probabilities of other three jumps are set to zero). 
\item check whether the new position of the particle leads to the better
optimum. If yes, change $i_b$ and $(x_b,y_b)$ accordingly.
\item do not move the best particle if not profitable.
\item when all particles from the swarm make their jumps, the iteration
is finished. Repeat it at a prescribed times or until some other stopping
criteria are met.
\end{itemize}

To test the algorithm performance, we tried it on some benchmark optimization
test functions. For each test function and for each number of iterations
one thousand independent numerical experiments were performed and the success
rate of the algorithm was calculated. The criterion of success was the 
following
\begin{eqnarray} &&
|x_f-x_m| \le \left \{ \begin{tabular}{l} $10^{-3}\,|x_m|,\;\; {\mathrm if} 
\;\; |x_m|> 10^{-3}$ \\  $10^{-3} ,\;\; {\mathrm if} \;\; |x_m|\le 10^{-3}$ 
\end{tabular},\right . \nonumber \\  && |y_f-y_m| \le \left \{ 
\begin{tabular}{l} $10^{-3}\, |y_m|,\;\; {\mathrm if} \;\; |y_m|> 10^{-3}$ 
\\  $10^{-3} ,\;\; {\mathrm if} \;\; |y_m|\le 10^{-3}$ \end{tabular}, \right . 
\label{eq10} \end{eqnarray}
where $(x_m,y_m)$ is the true position of the global optimum and $(x_f,y_f)$
is the  position found by the algorithm.
The results are given in the table \ref{tb1}.1. 
The test functions itself are defined in the appendix. Here we give only some 
comments about the algorithm performance. 
\begin{table}[ht]
\label{tb1}
\begin{center}
\caption{Success rate of the algorithm in percentages for various test
functions and for various numbers of iterations. Swarm size $n_p=20$.}
\begin{tabular}{|c|c|c|c|c|c|c|c|c|}
\hline
Function & \multicolumn{8}{c|}{Iterations} \\
\cline{2-9} 
Name & 50 & 100 & 200 & 300 & 400 & 500 & 600 & 700 \\ \hline \hline
Chichinadze & 35.5 & 97 & 100 & 100 & 100 & 100 & 100 & 100 \\ \hline
Schwefel & 99.4 & 99.5 & 99.8 & 99.3 & 99.2 & 99.8 & 100 & 99.6 \\ \hline
Ackley & 100 & 100 &  100 & 100 & 100 & 100 & 100 & 100 \\ \hline
Matyas & 88.9 & 100 & 100 & 100 & 100 & 100 & 100 & 100 \\ \hline
Booth & 100 & 100 & 100 & 100 & 100 & 100 & 100 & 100 \\ \hline
Easom & 93.6 & 100 & 100 & 100 & 100 & 100 & 100 & 100 \\ \hline
Levy5 & 98.4 & 99.5 & 99.4 &  99.3 &  99 & 99  & 99.1 & 99.5 \\ \hline
Goldstein-Price & 100 & 100 & 100 & 100 & 100 & 100 & 100 & 100 \\ \hline
Griewank & 76.3 & 99.7 & 100 & 100 & 100 & 100 & 100 & 100 \\ \hline
Rastrigin & 100 & 100 & 99.8 & 99.9 & 100 & 99.9 & 99.9 & 100 \\ \hline
Rosenbrock & 43.6 & 90.4 & 99.8 & 100 & 100 & 100 & 100 & 100 \\ \hline 
Leon & 13.8 & 52.1 & 82 & 91.6 & 97.6 & 99.1 & 99.6 & 99.8 \\ \hline
Giunta & 100 & 100 & 100 & 100 & 100 & 100 & 100 & 100 \\ \hline
Beale & 99.7 & 100 & 100 & 100 & 100 & 100 & 100 & 100 \\ \hline
Bukin2 & 61.8 & 84.4 & 93.8 & 97.8 & 98.6 & 99.3 & 99.7 & 99.8 \\ \hline
Bukin4 & 99.6 & 100 & 100 & 100 & 100 & 100 & 100 & 100 \\ \hline
Bukin6 & 0.2 & 0.1 & 0 & 0.2 & 0 & 0.1 & 0.2 & 0.1  \\ \hline
Styblinski-Tang & 100 & 100 & 100 & 100 & 100 & 100 & 100 & 100 \\ \hline
Zettl & 100 & 100 & 100 & 100 & 100 & 100 & 100 & 100 \\ \hline 
Three Hump Camel & 100 & 100 & 100 & 100 & 100 & 100 & 100 & 100 \\ \hline
Schaffer & 8.2 & 34.7 & 60.7 & 71.2 & 77.8 & 78.9 & 80.4 & 83.9 \\ \hline
Levy13 & 100 & 100 & 100 & 100 & 100 & 100 & 100 & 100 \\ \hline 
McCormic & 100 & 100 & 100 & 100 & 100 & 100 & 100 & 100 \\
\hline
\end{tabular}
\end{center}
\end{table}

For some test problems, such as Chichinadze, Ackley, Matyas, Booth, Easom,
Goldstein-Price, Griewank, Giunta, Beale, Bukin4, Styblinski-Tang, Zettl,
Levy13, McCormic and Three Hump Camel Back, the algorithm is triumphant. 

Matyas problem seems easy, because the function is only quadratic. However 
it is very flat near the line $x=y$ and this leads to problems for 
many global optimization algorithms. 

Easom function is a unimodal test function which is expected to be hard for 
any stochastic algorithms, because vicinity of its global minimum has a small 
area compared to the search space. Surprisingly our algorithm performs quite
well for this function and one needs only about 100 iterations to find the 
needle of the global minimum in a haystack of the search space.

Schwefel function is deceptive enough  to cause search algorithms to 
converge in the wrong direction. This happens because the global minimum is 
geometrically distant from the next best local minima. In some small fraction
of events our algorithm is also prone to converge in the wrong direction and
in these cases the performance seems not to improve by further increasing
the number of iterations. But the success rate is quite high. Therefore
in this case it is more sensible to have two or more independent tries
of the algorithm with rather small number of iterations each.  

Rastrigin function is a multimodal test function which have plenty of hills 
and valleys. Our algorithm performs even better for this function, but the 
success is not universal either. 

Rosenbrock function  is on contrary unimodal. Its minimum is situated in a 
banana shaped valley with a flat bottom and is not easy to find.
The algorithm needs more than 200 iterations to be successful 
in this case. Leon function is of the similar nature,
with even more flat bottom and the convergence in this case is 
correspondingly more slow.

Griewank, Levy5 and Levy13 are multimodal test functions. They are considered 
to be difficult for local optimizers because of the very rugged landscapes and
very large number of local optima. For example, Levy5 has 760 local minima in 
the search domain but only one global minimum and Levy13 has 900 local minima.
Test results reveal a small probability that our algorithm becomes stuck in 
one of the local minima for the Levy5 function.

Giunta function simulates the effects of numerical noise by means of a high
frequency, low amplitude sine wave, added to the main part of the function.
The algorithm is successful for this function.

Convergence of the algorithm is rather slow for Bukin2 function, and
especially for the Schaffer function. This latter problem is hard because
of the highly variable data surface features many circular local optima,
and our algorithm becomes, unfortunately, often stuck in the optima nearest
to the global one.

At last, the algorithm fails completely for the Bukin6 function. This function
has a long narrow valley which is readily identified by the algorithm. But the
function values differ very small along the valley. Besides the surface is
non-smooth in the valley with numerous pitfalls. This problem seems hopeless 
for any stochastic algorithm based heavily on random walks, because one has
to chance upon a very vicinity of the global optimum to be successful. The 
non-stochastic component of our algorithm (calculation of jump probabilities  
to mimic the quantum tunneling) turns out to be of little use for this 
particular problem.

\section{Concluding remarks}
The Quantum Swarm Optimization algorithm presented above emerged while trying
to generalize in the two-dimensional case a ``quantum mechanical'' algorithm
for automatic location of photopeaks in the one dimensional histograms 
\cite{1}.

`` Everything has been said before, but since nobody listens we have to keep 
going back and beginning all over again'' \cite{9}. After this investigation 
was almost finished, we discovered the paper \cite{10} by Xie, Zhang and Yang
with the similar idea to use the simulation of particle-wave duality in 
optimization problems. However their realization of the idea is quite 
different.

Even earlier, Levy and Montalvo used the tunneling method for global 
optimization \cite{11}, but without referring to quantum behavior. Their
method consisted in a transformation of the objective function, once
a local minimum has been reached, which destroys this local minimum
and allows to tunnel classically to another valley.

We found also that the similar ideology to mimic Nature's quantum behavior 
in optimization problems emerged in quantum chemistry and led to such 
algorithms as quantum annealing \cite{12} and Quantum Path Minimization 
\cite{13}.

Nevertheless, the Quantum Swarm Optimization is conceptually rather different
from these developments. We hope it is simple and effective enough
to find an ecological niche in a variety of global optimization algorithms.

\section*{Appendix}
Here we collect the test functions definitions, locations of their optimums 
and the boundaries of the search space.  The majority of them was
taken from \cite{14,15,16}, but we also provide the original reference when 
known.

Chichinadze function \cite{16,17}
$$F(x,y)=x^2-12x+11+10\cos{\frac{\pi}{2}x}+8\sin{(5\pi x)}-
\frac{1}{\sqrt{5}}\exp{\left (-\frac{(y-0.5)^2}{2}\right )},$$
$$-30\le x,y \le 30,\;\; F_{min}(x,y)=F(5.90133,0.5)=-43.3159.$$

Schweffel function \cite{18}
$$F(x,y)=-x\sin{\sqrt{|x|}}-y\sin{\sqrt{|y|}}, $$
$$-500\le x,y \le 500,\;\; F_{min}(x,y)=F(420.9687,420.9687)=-837.9658.$$

Ackley function  \cite{19}
$$F(x,y)=20[1-e^{-0.2\sqrt{0.5(x^2+y^2)}}]-
e^{0.5[\cos{(2\pi x)}+\cos{(2\pi y)}]}+e, $$
$$-35\le x,y \le 35,\;\; F_{min}(x,y)=F(0,0)=0.$$

Matyas function \cite{15}
$$F(x,y)=0.26 (x^2+y^2)-0.48 xy,$$
$$-10\le x,y \le 10,\;\; F_{min}(x,y)=F(0,0)=0.$$

Booth function \cite{16}
$$F(x,y)=(x+2y-7)^2+(2x+y-5)^2$$
$$-10\le x,y \le 10,\;\; F_{min}(x,y)=F(1,3)=0.$$

Easom function \cite{20}
$$F(x,y)=-\cos{x}\cos{y}\exp{[-(x-\pi)^2-(y-\pi)^2]},$$
$$-100\le x,y \le 100,\;\; F_{min}(x,y)=F(\pi,\pi)=-1.$$

Levy5 function \cite{15}
$$F(x,y)=\sum_{i=1}^5 i\cos{[(i-1)x+i]}\sum_{j=1}^5 j\cos{[(j+1)y+j]}+$$ 
$$+(x+1.42513)^2+(y+0.80032)^2,$$
$$-100\le x,y \le 100,\;\; F_{min}(x,y)=F(-1.30685,-1.424845)=-176.1375.$$

Goldstein-Price function \cite{15}
$$F(x,y)=\left [1+(x+y+1)^2 (19-14x+3x^2-14y+6xy+3y^2)\right ] \times $$
$$\times \left [30+(2x-3y)^2 (18-32x+12x^2+48y-36xy+27y^2)\right ], $$
$$-2\le x,y \le 2,\;\; F_{min}(x,y)=F(0,-1)=3.$$

Griewank function \cite{5,15}
$$F(x,y)=\frac{x^2+y^2}{200}-\cos{x}\cos{\frac{y}{\sqrt{2}}}+1,$$
$$-100\le x,y \le 100,\;\; F_{min}(x,y)=F(0,0)=0.$$

Rastrigin function \cite{21} 
$$F(x,y)=x^2+y^2-10\cos{(2\pi x)}-10\cos{(2\pi y)}+20,$$
$$-5.12\le x,y \le 5.12,\;\; F_{min}(x,y)=F(0,0)=0.$$

Rosenbrock function  \cite{15}
$$F(x,y)=100(y-x^2)^2+(1-x)^2,$$
$$-1.2\le x,y \le 1.2,\;\; F_{min}(x,y)=F(1,1)=0.$$

Leon function \cite{22}
$$F(x,y)=100(y-x^3)^2+(1-x)^2,$$
$$-1.2\le x,y \le 1.2,\;\; F_{min}(x,y)=F(1,1)=0.$$

Giunta function \cite{23}
$$F(x,y)=\sin{\left ( \frac{16}{15}x-1\right ) }+\sin^2{\left(\frac{16}{15}x
-1\right ) }+
\frac{1}{50}\sin{\left [ 4\left(\frac{16}{15}x-1\right ) \right ]}+$$ 
$$+\sin{\left ( \frac{16}{15}y-1\right ) }+\sin^2{\left ( \frac{16}{15}y-
1\right ) }+
\frac{1}{50}\sin{\left [ 4\left(\frac{16}{15}y-1\right)\right ]}+0.6,$$
$$-1\le x,y \le 1,\;\; F_{min}(x,y)=F(0.45834282,0.45834282)=0.0602472184$$

Beale function  \cite{15}
$$F(x,y)=(1.5-x+xy)^2+(2.25-x+xy^2)^2+(2.625-x+xy^3)^2,$$
$$-4.5\le x,y \le 4.5,\;\; F_{min}(x,y)=F(3,0)=0.$$

Bukin2 function \cite{24}
$$F(x,y)=100(y-0.01x^2+1)+0.01(x+10)^2,$$
$$-15\le x \le -5,\;-3\le y \le 3,\;\; F_{min}(x,y)=F(-10,0)=0.$$

Bukin4 function \cite{24}
$$F(x,y)=100y^2+0.01|x+10|,$$
$$-15\le x \le -5,\;-3\le y \le 3,\;\; F_{min}(x,y)=F(-10,0)=0.$$

Bukin6 function \cite{24}
$$F(x,y)=100\sqrt{|y-0.01x^2|}+0.01|x+10|,$$
$$-15\le x \le -5,\;-3\le y \le 3,\;\; F_{min}(x,y)=F(-10,1)=0.$$

Styblinski-Tang function \cite{25}
$$F(x,y)=\frac{1}{2}\left [x^4-16x^2+5x+y^4-16y^2+5y\right ],$$
$$-5\le x,y \le 15,\;\; F_{min}(x,y)=F(-2.903534,-2.903534)=-78.332.$$

Zettl function \cite{22}
$$F(x,y)=(x^2+y^2-2x)^2+0.25x,$$
$$-5\le x,y \le 5,\;\; F_{min}(x,y)=F(-0.0299,0)=-0.003791.$$

Three Hump Camel back function \cite{14}
$$F(x,y)=2x^2-1.05x^4+\frac{x^6}{6}+xy+y^2,$$
$$-5\le x,y \le 5,\;\; F_{min}(x,y)=F(0,0)=0.$$

Schaffer function \cite{26} 
$$F(x,y)=0.5+\frac{\sin{\sqrt{x^2+y^2}}-0.5}{[1+0.001(x^2+y^2)]^2},$$
$$-100\le x,y \le 100,\;\; F_{min}(x,y)=F(0,0)=0.$$

Levy13 function \cite{14}
$$F(x,y)=\sin^2{(3\pi x)}+(x-1)^2\left [ 1+\sin^2{(3\pi y)}\right ]+
(y-1)^2\left [1+\sin^2{(2\pi y)} \right ], $$ 
$$-10\le x,y \le 10,\;\; F_{min}(x,y)=F(1,1)=0.$$

McCormic function \cite{27}
$$F(x,y)=\sin{(x+y)}+(x-y)^2-1.5x+2.5y+1,$$
$$-1.5\le x \le 4,\; -3\le x \le 4 \;\; F_{min}(x,y)=F(-0.54719,-1.54719)=
-1.9133.$$

\section*{Acknowledgments}
Support from the INTAS grant No. 00-00679 is acknowledged.

\end{document}